\documentclass[aps,prb,reprint,superscriptaddress]{revtex4-2}
\usepackage{bm}
\usepackage{braket}
\usepackage{mathtools}
\usepackage{color}
\usepackage{comment}
\usepackage{physics}
\usepackage{amsmath,amssymb,amsfonts} 
\usepackage{here}
\usepackage{url}
\usepackage[colorlinks=true,citecolor=blue,linkcolor=blue,filecolor=blue,urlcolor=blue]{hyperref} 
 \usepackage{graphicx} 

\begin{document}
\title{Theory of clusterization in orbitally degenerate transition-metal compounds driven by lattice instabilities
 }
\author{Soshun Ozaki}
 \email{soshunozaki@gmail.com}
\affiliation{Department of Basic Science, University of Tokyo, Meguro, Tokyo 153-0041, Japan}
\affiliation{Department of Physics, Chuo University, Bunkyo, Tokyo 112-8551, Japan}
\affiliation{Max Planck Institute for Solid State Research, Heisenbergstrasse 1, D-70569 Stuttgart, Germany}
\author{Kota Mitsumoto}
\affiliation{Department of Basic Science, University of Tokyo, Meguro, Tokyo 153-0041, Japan}
\author{Chisa Hotta}
 \email{chisa@phys.c.u-tokyo.ac.jp}
\affiliation{Department of Basic Science, University of Tokyo, Meguro, Tokyo 153-0041, Japan}

\date{\today}

\begin{abstract}
We derive an effective orbital-lattice model with quantum $S=1$ degrees of freedom for transition-metal compounds, 
providing a microscopic understanding of cluster formation driven by the cooperative interplay of 
spin, orbital, and lattice degrees of freedom.
Motivated by the trimerized phases observed in LiVS$_2$ and LiVO$_2$, 
we consider a triangular-lattice three-orbital system with two electrons per site occupying the threefold-degenerate $t_{2g}$ manifold. 
Starting from a multiorbital Kanamori-Hubbard Hamiltonian, we project the low-energy sector onto the local $S=1$ triplet manifold, in which two electrons occupy different orbitals according to Hund's coupling. 
The resulting effective model exhibits exchange networks whose geometry is determined by the orbital configuration. 
However, the orbital-driven exchange interactions alone do not stabilize the experimentally observed trimer phase. 
We find that by incorporating ionic lattice displacements that modulate transfer integrals and induce bond-dependent exchange couplings on shortened and elongated bonds, the phase competition is qualitatively altered, 
leading to the robust stabilization of a trimerized ground state within a fully quantum-mechanical framework. 
We further show that a simplified orbital-lattice model, in which the spin-exchange energy is replaced by effective bond energies, 
faithfully reproduces the essential ground-state properties of the microscopic model. 
This reduced description enables large-scale finite-temperature simulations and reveals a rich sequence of thermal phase transitions, including first-order, second-order, and Kosterlitz-Thouless transitions into distinct spin-, orbital-, and lattice-ordered phases. 
\end{abstract}
\maketitle

\section{Introduction}
The interplay of multiple degrees of freedom in solids gives rise to a rich variety of magnetic, 
dielectric, and electronic phenomena, particularly in transition-metal compounds \cite{Khomskii2021ChemRev}.
A key ingredients is the orbital degeneracy of partially filled $t_{2g}$ shells in a crystal field from the surrounding ligands, which couples strongly to spin, charge, and lattice degrees of freedom. 
The competition and cooperation among these degrees of freedom often drive structural instabilities accompanied by orbital ordering, reflecting a delicate balance between electronic correlations, magnetism, and lattice distortions.
Indeed, such structural responses frequently serve as signatures of
complex collective behavior in correlated materials, including magnetoelectricity \cite{Yamauchi2012PRB,Kocsis2023PRL}, colossal magnetoresistance \cite{Feiner1999PRB}, orbital-glass formation \cite{Fichtl2005PRL,Mitsumoto2020PRL}, and valence-bond crystallization \cite{Katayama2024,Okamoto2008PRL,Shiomi2022PRB}.
\par
Among the various forms of structural organization, a particularly striking phenomenon 
is the spontaneous formation of clusters of transition-metal ions. 
Structural studies have revealed a variety of cluster patterns, 
including octamers in CuIr$_2$S$_4$ \cite{Furubayashi1994JPSJ,Radaelli2002Nature}, 
helical dimers in MgTi$_2$O$_4$ \cite{Isobe2002JPSJ,Schmidt2004PRL,Torigoe2018PRB}, 
heptamers in AlV$_2$O$_4$ \cite{Horibe2006PRL}, later interpreted as combinations of trimers and tetramers \cite{Browne2017PRMaterials},
as well as a trimeron formation in Fe$_3$O$_4$ \cite{Senn2012Nature}. 
These clusterizations are discussed theoretically in many different contexts 
including orbital ordering, spin-orbital coupling, 
and valence-bond dimer formations
\cite{Pen1997PRL, Feiner1997PRL, DiMatteo2004PRL, DiMatteo2005PRB, Khomskii2005PRL, Khomskii2005PTPS, Oles2006PRL, Jackeli2008PRL}. 
Recently, another trimer formation is found in pyrochlore oxide, CsW$_2$O$_6$\cite{Okamoto2020Ncomm}, 
relevant to the lattice instability induced by the underlying flat band singularity
\cite{Nakai2022Ncomm}. 
\par
Among these, the layered compounds with a triangular lattice, LiV$X_2$, 
provide a prototypical platform to study such phenomena
\cite{Onoda1991JPSJ, Tian2006PhysicaB, Kojima2019PRB, Tanaka2009JPSJ, Kawasaki2011JPCS, Katayama2021npjQM, Kojima2023PRB, Jinno2013PRB, Katayama2009PRL}, 
where each V ion hosts two electrons in the triply degenerate $t_{2g}$ orbitals 
in an approximately octahedral ligand field. 
Indeed, the trimerization-related phenomena have been widely observed in other 
materials having transition metal ions with partially filled $t_{2g}$ orbitals, 
such as NaVO$_2$\cite{McQueen2008PRL, Jia2009PRB}, and titanium halides\cite{Hnni2017}. 
However, the nature of trimerization has remained elusive despite decades of intensive investigation 
since Goodenough proposed a trimerized state based on the vanadium lattice instability, 
attributing the nonmagnetic behavior to molecular orbital formation within trimers that quenches local moments 
\cite{Goodenough1991PRB}. 
The nonmagnetic molecular orbital picture was supported by the X-ray diffraction 
and extended X-ray absorption fine structure measurements \cite{Katayama2009PRL,Kojima2019PRB}, 
as well as NMR experiments indicating symmetry lowering below the transition temperature. 
However, the persistence of local $S=1$ moments observed experimentally is difficult to reconcile with 
a simple molecular-orbital picture, 
and several more intricate microscopic mechanisms have been proposed\cite{Tanaka2009JPSJ,Kawasaki2011JPCS}. 
\par
From a theoretical perspective, stabilizing trimer states under the competing spin and orbital degrees of freedom has proven to be highly nontrivial. 
Starting from the Kanamori Hamiltonian as a minimal microscopic model for $t_{2g}$ systems, 
early work by Pen \textit{et al.} \cite{Pen1997PRL} proposed a picture 
that the local $S=1$ moments emerge from Hund’s coupling, which forms a trimer singlet due to orbital-selective exchange couplings. 
However, this scenario was based on three-site exact diagonalization (ED) 
and did not fully establish the energetic stability of the trimer singlet. 
A subsequent study \cite{Yoshitake2011JPSJ} derived Kugel–Khomskii-type effective models and found that, within a classical treatment of $S=1$ spins with orbital degeneracy, the trimer state is generally unstable against ferrimagnetic ordering. 
Within their approach, the orbital trimerization was only possible by 
an additional crystal field splitting that lifts the $t_{2g}$ degeneracy into $a_{1g}$ and $e_g'$ levels, 
and drives the formation of the $S=1/2$ singlet pairs on each trimerized bond, 
thus does not conform to the $S=1$ picture. 
\par
These theoretical difficulties stand in sharp contrast to the robust experimental evidence for 
the compatible observations of the $S=1$ moments and the trimer ordering in this class of materials, 
highlighting a fundamental gap in our understanding.
In this work, we address this issue by constructing an effective model for trimer formation that explicitly incorporates spin and orbital 
degrees of freedom, guided by experimentally observed lattice distortions.
Starting from a Kanamori-type Hamiltonian, similarly to other works, we derive a low-energy description, 
where the additional lattice-displacement degrees of freedom naturally 
captures the stabilization of trimer states. 
\par
The remainder of this paper is organized as follows. 
In Sec.~II, we derive the effective orbital-spin-lattice model from the Kanamori Hamiltonian by incorporating orbital configurations and lattice displacements particularly relevant to LiVO$_2$ and LiVS$_2$. 
We further simplify it to the orbital-lattice model by replacing the quantum $S=1$ spin exchange interactions 
with constant bond energies. 
In Sec.~III, we clarify the ground state phase diagram of both the orbital-spin-lattice model and the orbital-lattice model, 
and obtain a finite-temperature phase diagram of the latter case, where we rely on the classical Monte Carlo simulation. 
As we summarize in Sec.~IV, the two models consistently explain the low-energy properties of the triangular-lattice-based transition metal compounds, and naturally realize the $S=1$ trimer formation.

\begin{figure*}
    \centering
    \includegraphics[width=\linewidth]{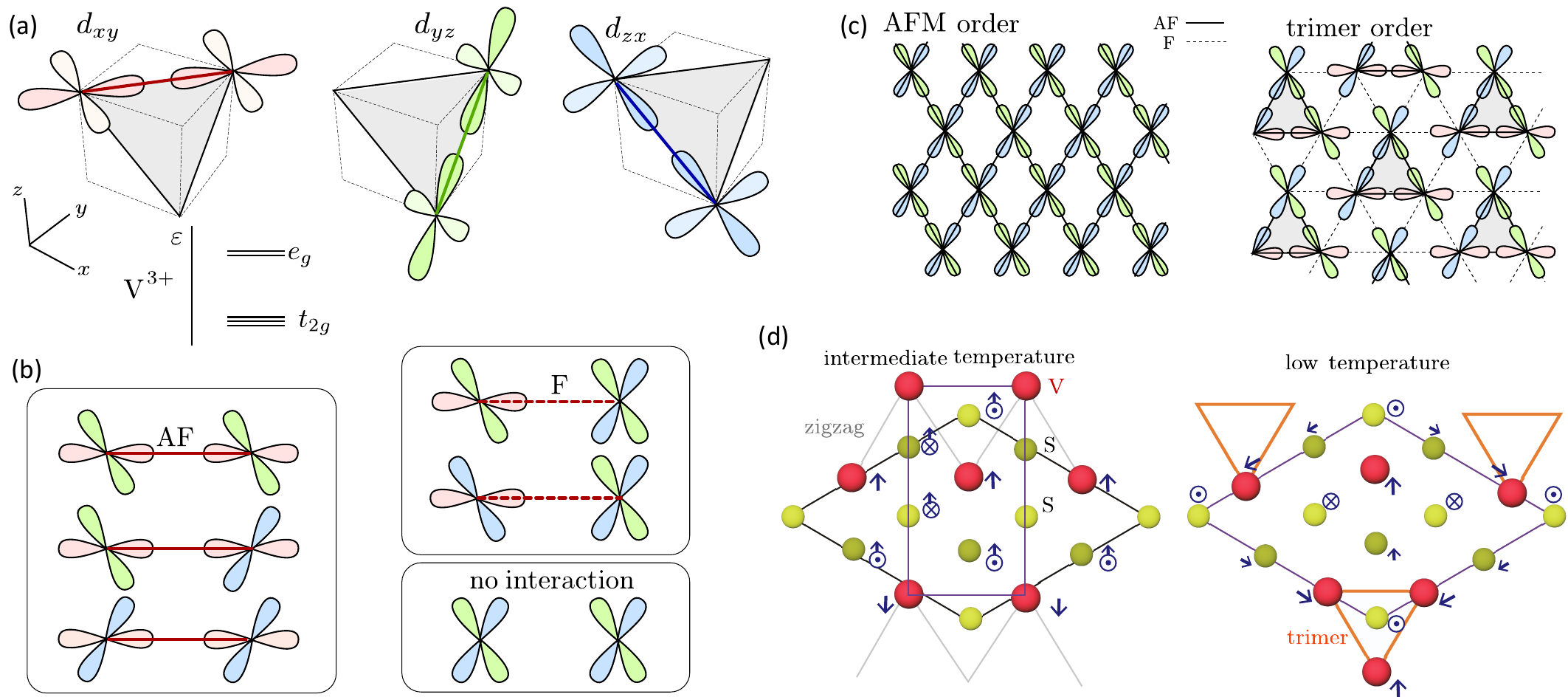}
    \caption{(a) The spatial configurations of $t_{2g}$ orbitals on a triangular lattice,
    where we set the triangular plane perpendicular to the $[1 1 1]$ direction in the $xyz$-coordinate of a crystal field.
    Three panels show the bond direction along which $d_{xy}, d_{yz}, d_{zx}$
    orbitals have the $\sigma$ type of overlaps where we expect finite $t$.
    We set the other two bond directions to have no overlap.
    (b) Configurations of two of the three orbitals which the electrons occupy on the neighboring two sites along the $xy$ bond, where, unlike panel (a) we only show half of the component of each orbital that lies on the triangular plane.
     The $S=1/2$'s on the two orbitals form $S=1$,
     and when the $d_{xy}$ on both sides are occupied, they have antiferromagnetic exchange (solid lines), whereas if one $d_{xy}$ is occupied and the other is empty, we have ferromagnetic exchange (broken lines).
    When both $d_{xy}$'s are empty, there is no spin exchange.
    (c) Two different regular configurations of orbitals forming square lattice bonds with antiferromagnetic exchange,
    and trimer order with intra-trimer antiferromagnetic exchange and inter-trimer ferromagnetic exchange.
(d) Lattice displacements of LiVS$_2$ observed experimentally for the 
intermediate zigzag state and the low-temperature trimer state\cite{Katayama2021npjQM}, 
where the arrows roughly indicate the degree of displacements.
In both phases, V$^{3+}$ ions move inside and out of the plane 
depending on their positions.
}
    \label{f1}
\end{figure*}

\section{Deriving an effective spin-lattice model}
\subsection{Kanamori Hamiltonian}
Let us consider a triangular lattice in a unit of $3d$ ions, with in mind the V$^{3+}$ ions of LiVS$_2$ and LiVO$_2$.
The $3d$ states split into doubly degenerate $e_g$ and triply degenerate $t_{2g}$ levels
in a ligand field of approximate $O_H$ symmetry, 
and among the six $t_{2g}$ orbitals including spin degeneracy, two are occupied.
\par
The starting point is the Kanamori-type model, also referred to as a multiorbital Hubbard model, common to 
the related studies\cite{Pen1997PRL,Yoshitake2011JPSJ}, given as 
\begin{align}
    {\cal H}&= {\cal H}_{U}+ {\cal H}_t,\notag \\
 {\cal H}_{U} &=
    +\sum_{i,\alpha,\beta} (U'+2J_H \delta_{\alpha \beta}) n_{i\alpha \uparrow} n_{i\beta \downarrow} \nonumber \\
    &+\frac{J_H}{2}\sum_{i,\alpha\neq \beta, \sigma,\sigma'}
    (c_{i\alpha \sigma}^\dagger c_{i\beta\bar{\sigma}}^\dagger c_{i\alpha \bar{\sigma}} c_{i\beta \sigma} + c_{i\alpha \sigma}^\dagger c_{i\alpha\bar{\sigma}}^\dagger c_{i\beta \bar{\sigma}} c_{i\beta \sigma}),
\notag\\
{\cal H}_t&= -\sum_{\langle i,j\rangle,\alpha,\sigma}
    t_{\alpha}^{ij}c_{i\alpha\sigma}^\dagger c_{j\alpha\sigma} + {\rm H.c.},
    \label{eq:hubbard}
\end{align}
where $c^{(\dagger)}_{i\alpha\sigma}$ annihilates (creates) an electron in orbital $\alpha(=d_{xy},d_{yz},d_{zx})$ with spin $\sigma$ at site $i$, and $n_{i\alpha \sigma}$ denotes the corresponding number operator.
The on-site term, ${\cal H}_{U}$, includes the intra-orbital and inter-orbital Coulomb interactions $U$ and $U'$,
and the Hund's coupling, $J_H$, where we adopt the relationship, $U\equiv U'+2J_H$, that applies for
the octahedral crystal field. 
\par
The interesting feature of Eq.(\ref{eq:hubbard}) for the triangular lattice system is that 
the inter-site hopping term has a strong directional dependence, as shown schematically in Fig.~\ref{f1}(a); 
Each of the three degenerate orbitals extends along one of the three bond directions of a triangular lattice,
and the $\sigma$ bonding overlap occurs only along that direction, e.g. $d_{xy}$ has an overlap along $[110]$ bond 
and are negligible along the other two directions of the triangle. 
This anisotropic hopping is denoted as $t_{\alpha}^{ij}$, and is set to be nonzero only when the orbital $\alpha$ coincides with the index mediating a $\sigma$ bond between sites $i$ and $j$. 
\par
Let us first outline the previous arguments given on this Hamiltonian, with particular focus on the singlet trimer formation. 
Pen {\it et al.} \cite{Pen1997PRL} considered the $S=1$ high spin state on each site realized by the Hund's coupling,
where there are $\,_{3}C_{2}$ choices of electronic configuration of selecting the orbitals.
Depending on these choices, the exchange energy between the neighboring $S=1$'s is estimated up 
to second order in terms of $t$, 
as shown schematically in Fig.~\ref{f1}(b);
If the neighboring sites are aligned along the $xy$ direction and $d_{xy}$ are occupied on both,
the energy gain measured against the ferromagnetic (up-up) configuration
is given as $-2t^2/U \equiv -2J_0$ from the second order perturbation process.
If only one of the two sites has occupancy in $d_{xy}$,
that occupied electron can hop to the vacant site and come back, which gives the energy gain of
$-t^2/U=-J_0$, independent of the spin orientation.
One can thus assign the orbital configuration on each site and sum these energy gains per bond;
For the antiferromagnetic (AFM) ordering shown in the left panel of Fig.~\ref{f1}(c),
the AFM bonds form a square lattice, and the sum of exchange couplings amounts to $E_{\rm AF}=-2J_0 \times 2N$.
For the trimer configuration in the right panel of Fig.~\ref{f1}(c), the intra-trimer AFM couplings and
inter-trimer FM couplings yield, $E_{\rm tri}=-2J_0\times N+ 2(-J_0)\times N$.
As these two estimates are equal, one needs to take further account of the quantum many-body effect.
The authors performed the three-site ED of Eq.(\ref{eq:hubbard}) and showed that the state with total spin $S^{\rm tot}=0$ can lower the energy against $S^{\rm tot}=1$ at $t\gtrsim 0.1 U$ with $J_H\simeq 0.16 U$.
\par
Later on, Yoshitake {\it et al.} \cite{Yoshitake2011JPSJ} adopted the same orbital-selective directional spin exchange coupling for $S=1$,
where they treated the spins classically by the Monte Carlo simulation,
and showed that the trimer configuration is energetically favorable against the AFM ordering,
whereas there exists a ferrimagnetic ordering of a larger unit cell, always lower in energy.
This may suggest that the orbital-selective $S=1$ model cannot afford the trimer ordering.
Therefore, they discarded this model and instead 
introduced the trigonal distortion of the crystal field,
which splits the degenerate $t_{2g}$ levels into an $e_g'$ doublet and an $a_{1g}$ singlet. 
Although this mechanism can stabilize a nonmagnetic trimer state, it no longer describes the trimer singlet state 
as a combination of three $S=1$'s. 
This is difficult to reconcile with the experimental observations of LiVO$_2$, where the magnetic susceptibility and NMR measurements above the transition temperature were explained in terms of the localized $S=1$ moments of V$^{3+}$ ions \cite{Onoda1991JPSJ,Kawasaki2011JPCS}.
\par
These inconsistencies motivate us to consider the effect of spatially modulated lattice distortions on top of the orbital 
and spin degrees of freedom, that can give natural selection to the trimer ordering.
Indeed, recent experimental studies on LiVS$_2$
show that the trimerization is accompanied by the structural transition at
314 K into $P31m$, and above that temperature, there is another structure,
$P\bar 3m1$, which supports the zigzag type of displacement structure
with dynamical domains of $\sim$ 100 \AA\ scale stabilized within the temperature range of 314 K $< T \lesssim 350$ K \cite{Katayama2021npjQM}.
The lattice displacement patterns obtained from the structural analysis are
illustrated in Fig.~\ref{f1}(d)
for the low-temperature trimer phase and the intermediate zigzag phase.

\begin{figure}
    \centering
    \includegraphics[width=\linewidth]{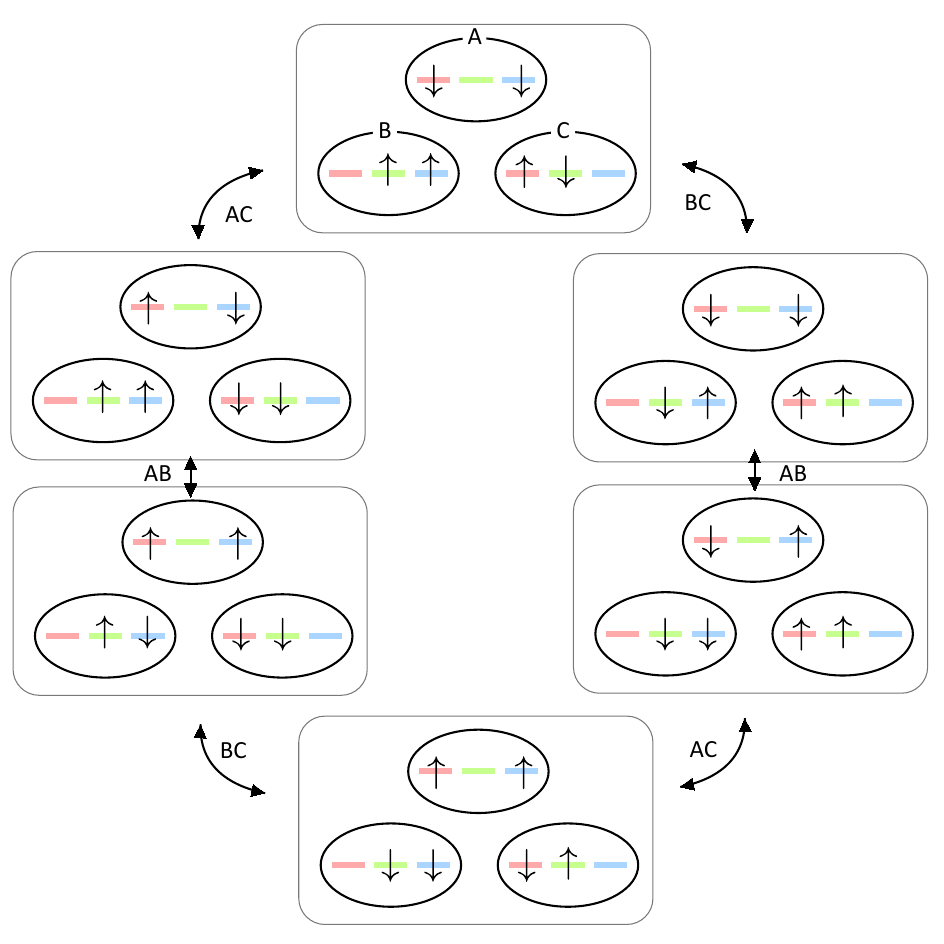}
    \caption{Schematic illustration of the major component of the
wave function of the $S^{\rm tot}=0$ lowest energy state realized by ED on a three-site cluster
for Eq.(\ref{eq:hubbard}).
}
    \label{f2}
\end{figure}
\subsection{Revisiting the exact $S=1$ trimerized state}
We first clarify what kind of microscopic configuration is present in the $S_{\rm tot}=0$ singlet state
of the three-site trimer in terms of ED on Eq.(\ref{eq:hubbard}),
as the earlier study has focused only on the total spin and not on how it is stabilized
under the quantum many-body effect.
In the case of LiVO$_2$, model parameters for Eq.(\ref{eq:hubbard}) are $t\sim 1$eV with $U=5.1$ eV, $J_H=0.8$eV.
We find a unique lowest-energy eigenstate, $|\psi_{\rm Hubbard}(N=3)\rangle$, with $S_{\rm tot}=0$ at $N=3$, 
which consists of six different major electron configurations. 
The schematic illustration of these configurations is shown in Fig.~\ref{f2}, 
where the site indices A, B, and C are assigned to the three sites of the trimer.
The electron configurations are always fixed for all three sites because of the strong Coulomb repulsion and Hund's coupling. 
In this configuration, for example, the $d_{yz}$ is the only orbital where an electron is allowed to hop between B and C.
\par
This result is understood by a purely quantum $S=1$ Heisenberg model forming a triangle given as, 
\begin{align}
    H_{\rm 3site} &= J_{\rm AF} ({\bm S}_{\rm A} \cdot {\bm S}_{\rm B} + {\bm S}_{\rm B} \cdot {\bm S}_{\rm C} + {\bm S}_{\rm C} \cdot {\bm S}_{\rm A}), \label{eq:3site_spin}
\end{align}
where $\bm{S}_{\rm A}$, $\bm{S}_{\rm B}$, and $\bm{S}_{\rm C}$ are local $S=1$ spin operators at sites 
A, B, and C, respectively. 
This Hamiltonian is rewritten as $J_{\rm AF} (\bm S_{\rm tot}^2 - 6)/2$, 
and $S_{\rm tot}=0$ state becomes the unique ground state with energy $E_{\rm 3site}=-3J_{\rm AF}$, which is given explicitly as 
\begin{align}
\ket{\text{3s}}= &\frac{1}{\sqrt{6}} \big( \ket{1,-1,0} + \ket{-1,0,1} + \ket{0,1,-1} \nonumber \\
&- \ket{-1,1,0} - \ket{1,0,-1} - \ket{0,-1,1} \big),
\label{eq:3singlet}
\end{align}
spanned by the combination of states indexed by the $z$ component of spins, $|S_{\rm A}^z,S_{\rm B}^z,S_{\rm C}^z\rangle$. 
This ground state is equivalent to the one shown in Fig.~\ref{f2} for Eq.(\ref{eq:hubbard}), 
as we fix the orbital configuration, and the remaining degrees of freedom are the three $S=1$'s. 
The fidelity between the lowest $S_{\rm tot}=0$ state of the Hubbard model and the three-site singlet state is 
$\langle \psi_{\rm Hubbard}(N=3)|{\rm 3s}\rangle=0.792$. 
To be more precise, $J_{\rm AF}$ in Eq.(\ref{eq:3site_spin}) is derived from the fixed orbital configurations 
in Fig.~\ref{f2} perturbatively at the second order  in terms of $t$ 
as $J_{\rm AF}=J_0(1+\eta)/(1+2\eta)$ with  $\eta=J_H/U$ and $J_0=t^2/U$. 

\begin{figure}
    \centering
    \includegraphics[width=\linewidth]{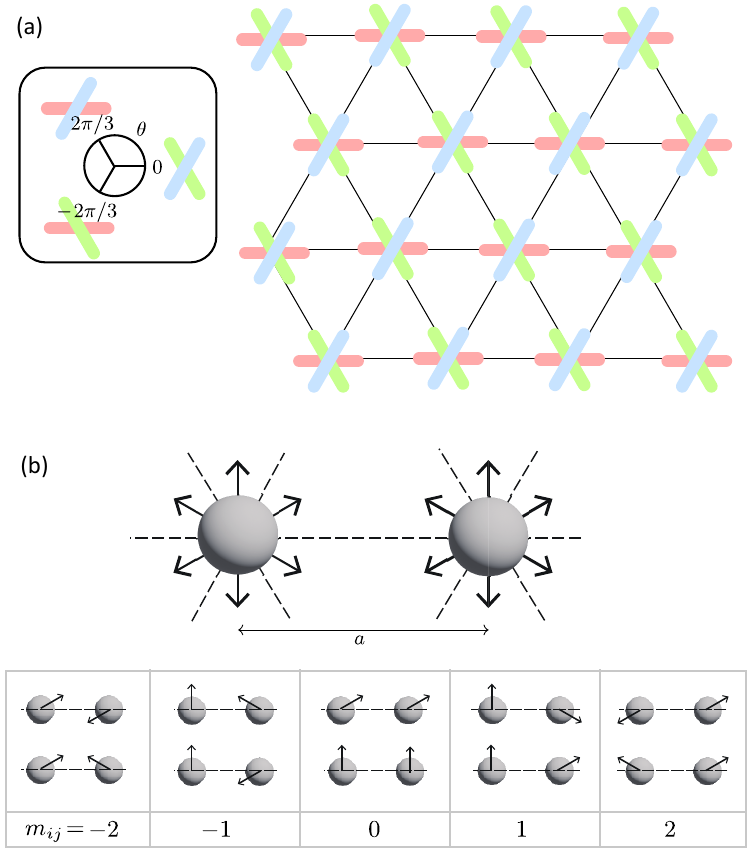}
\caption{(a) Schematic illustration of the effective Hamiltonian, Eq.(\ref{eq:effectivemodel}).
(b) Lattice displacement degrees of freedom of the adjacent two sites, whose directions are given
in arrows. The relative angles between the two displacements yield the degree of modulation of $J_{\rm F0}$, $J_{\rm F}$, and $J_{\rm AF}$
in the effective model of Eq.(\ref{eq:effectivemodel}).
}
    \label{f3}
\end{figure}
%
\begin{figure*}
    \centering
    \includegraphics[width=\linewidth]{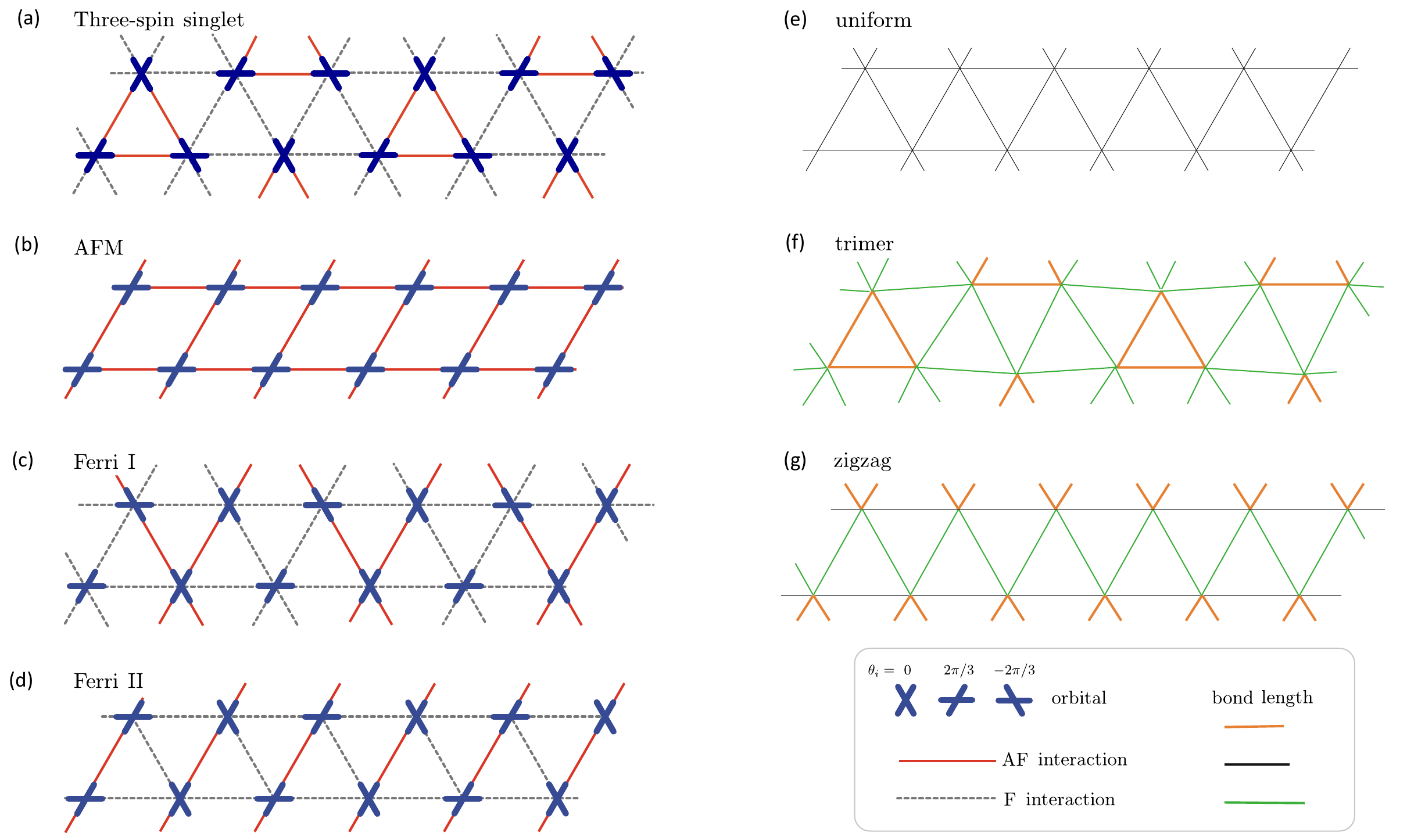}
    \caption{
    (a)--(d): Typical orbital configurations proposed by Yoshitake \textit{et al.}:
    (a) three-spin singlet phase, (b) AFM phase, (c) Ferri I phase, and (d) Ferri II phase.
    Each vertex corresponds to a vanadium atom. Blue thick short bars indicate the direction of electron propagation. Red and blue lines represent $S=1$ AFM and FM
    interactions, respectively.
    (e)--(g): Lattice distortion patterns experimentally observed in LiVS$_2$: (e) uniform phase, (f) trimer phase, and (g) zigzag phase. Black, yellow, and green
    lines represent bonds of natural length, compressed bonds, and stretched bonds,
    respectively.
    }
    \label{f4}
\end{figure*}
\subsection{Orbital-spin model for the bulk system}
We now derive the perturbative Hamiltonian up to second order in $t/U$ for the full triangular lattice 
by restricting the local Hilbert space to the Hund-coupled $S=1$ sector, 
and obtain the orbital-spin Hamiltonian~\cite{Yoshitake2011JPSJ},
\begin{align}
    {\cal H}_{\rm os}&=\sum_{\langle i,j \rangle} [\;
   \hat h_1 (n_{i,\alpha}\bar{n}_{j,\alpha}+ \bar{n}_{i,\alpha}n_{j,\alpha}) 
   + \hat h_2 n_{i,\alpha} n_{j,\alpha} \;] ,  \notag \\
   & \hat h_1 =-J_F {\bm S}_i\cdot {\bm S}_j -J_{F0}, \notag \\
   & \hat h_2 =J_{AF}({\bm S}_i\cdot {\bm S}_j-1), 
    \label{eq:effectivemodel}
\end{align}
where  ${\bm S_i}$ denotes $S=1$ spin operator at site $i$, 
the orbital index $\alpha$ has the same label as the $\langle i,j\rangle$ bond, 
and $\bar{n}_{i,\alpha}=1-n_{i,\alpha}$ is the hole occupation.
The exchange couplings are given by
\begin{align}
    J_{F0}&=\frac{1-\eta}{1-3\eta}J_0, \quad
    J_F=\frac{\eta}{1-3\eta} J_0, \notag\\
    J_{AF}&=\frac{1+\eta}{1+2\eta} J_0,
\end{align}
with $J_0=t^2/U$, $\eta=J_H/U$.
We now rephrase the physical implication of these terms using Figs.~\ref{f1}(b) and \ref{f3}(a).
When the $\langle i,j\rangle$ bond is pointing in the $xy$ direction,
and site $i$ and $j$ both have the $d_{xy}$ orbitals occupied,
we find $h_2\ne 0$ that provides an AFM coupling term,
generated by the kinetic exchange between two electrons induced by $t$.
Whereas if one of the $d_{xy}$ orbitals is occupied and the other is empty,
the corresponding orbital component of $\hat h_1$ is nonzero due to the second-order perturbation process in $\mathcal{H}_t$, yielding a double exchange-like FM term.
If both $d_{xy}$'s are empty, there is no room for an exchange interaction. 
\par
This Eq.(\ref{eq:effectivemodel}) is a quantum $S=1$ model coupled to orbital degrees of freedom,
spanned by $3\times 3$ local basis per site, 
and basically has the same form as the one discussed previously\cite{Pen1997PRL,Yoshitake2011JPSJ}. 
However, Ref.\cite{Yoshitake2011JPSJ} has treated $S=1$ as classical vector spins,
and Ref.\cite{Pen1997PRL} expanded their discussion for cases 
that the spin exchange operators, $(J_{F0}+J_F {\bm S}_i\cdot {\bm S}_j)$ and $J_{AF}(-1+{\bm S}_i\cdot {\bm S}_j)$, 
are replaced by the constant energy values, $-J_0$ and $-2J_0$, respectively. 
We perform a full quantum mechanical treatment on the $S=1$ part shortly in Sec.III and examine the consistency with 
these classical treatments. 
\par
We present typical orbital and corresponding bond configurations in Figs.~\ref{f4} (a)-(d).
The three-spin singlet state is characterized by the periodic orbital configurations
in the unit of three sites, that form AFM triangles, and the rest of the interactions are FM.
The AFM state forms a square lattice with uniform AF interactions.
The two ferrimagnetic states (Ferri I, Ferri II) are the ones introduced by the Monte Carlo study in Ref.\cite{Yoshitake2011JPSJ},
which constitute the ground state of the classical version of Eq.(\ref{eq:effectivemodel})
in the absence of the lattice displacement and trigonal crystal field. 

\subsection{Lattice displacements}
As we showed in Fig.~\ref{f1}(d), both zigzag and trimer types of lattice displacements are observed in LiVS$_2$.
We thus incorporate atomic displacements into the model.
When the lattice time-scale is much longer than that of the spins, the effect of lattice
distortion can be treated as a modulation of the exchange integrals,
reflecting the changes in the hopping amplitude $t$, together with an elastic energy cost.
We assume that the V atom moves in one of six discrete directions with a fixed
displacement length, as illustrated in Fig.~\ref{f3}(b),
where $a$ denotes the lattice constant of the undistorted triangular lattice.
Introducing a unit displacement parameter $\delta$, 
the relative distance between the adjacent $i$th and $j$th sites is given at a linear order in $\delta$ as 
\begin{equation}
    a \to (1+m_{ij} \delta)a \qquad (m_{ij}=-2,-1,0,1,2). 
\end{equation}
We show in Fig.~\ref{f3}(b) the regular displacement patterns for five classes of $m_{ij}$,
where totally 36 types of configurations are classified.
The upper and lower panel ones slightly differ in distance by $O(\delta^2)$ which are neglected.
The corresponding modulation of the hopping integral is given as 
\begin{equation}
    t\to (1-\frac{\nu}{2} m_{ij})t,
\end{equation}
where $\nu$ represents the magnetoelastic coupling, proportional to $\delta$. 
The exchange coupling is modified as 
\begin{equation}
    J_{\rm F/AF}\to J_{\rm F/AF}(1-\frac{\nu}{2} m_{ij})^2. \label{eq:jsubstitution}
\end{equation}
There is an associated elastic energy cost of $\kappa m_{ij}^2$ with the elastic constant $\kappa$,
leading to the lattice Hamiltonian, $H_{\rm lattice}=\sum_{\langle i,j \rangle}\kappa m_{ij}^2$.
The uniform, trimer, and zigzag types of periodic lattice modulations observed experimentally 
are shown in Fig.~\ref{f4} (e)--(g). 
The trimer includes two different types of $m_{ij}$ and the zigzag has three. 
%
\begin{table*}
    \caption{Numbers of uniform (undistorted), short, and long bonds
    under Trimer and Zigzag type of lattice distortions
    with AFM and FM couplings, assuming four different types of orbital configurations
    shown in Fig.~\ref{f4}. \label{table:bondcounts}}
    \begin{ruledtabular}
    \begin{tabular}{cc|ccccccc}
    Distortion & spin config. & Sites/unit &  uniform AFM  & long AFM & short AFM & uniform FM & long FM & short FM  \\ \hline
    Trimer & three-spin singlet  & 6 & 0 & 0  &6 &0& 12 & 0\\
    Trimer & AFM  & 6&0&9&3&0&0&0 \\
    Trimer & Ferri I \& II  & 12&0&8&4&0&16 & 8 \\
    Zigzag & three-spin singlet  & 6 & 2 & 2 & 2 & 4 & 4 & 4\\
    Zigzag & AFM  & 2 &2 & 1 &1&0&0&0 \\
    Zigzag & Ferri I \& II  & 4&0&2&2& 4& 2 & 2 \\
    \end{tabular}
    \end{ruledtabular}
\end{table*}
\subsection{Orbital-spin-lattice model and orbital-lattice model}
We now have the orbital-spin-lattice model given as a combination 
of Eqs.(\ref{eq:effectivemodel}) and (\ref{eq:jsubstitution}).
Let us then transform it to a simpler representation as
\begin{align}
    {\cal H}_{\rm eff}&= \sum_{\langle ij \rangle}
    \big[\:\hat h(\theta_i,\theta_j)(1-\frac{\nu}{2} m_{ij})^2 + \kappa m_{ij}^2\ \big] \notag \\
    & \hat h(\theta_i,\theta_j) = \hat h_\ell, \quad
     \ell= \delta_{\theta_i b_{ij}} + \delta_{\theta_j,b_{ij}}, 
    \label{eq:finalmodel}
\end{align}
with $\hat h_\ell$ given in Eq.(\ref{eq:effectivemodel}). 
Here, we introduce the parameter $b_{ij}=0,2\pi/3,-2\pi/3$, 
denoting the bond direction that has finite hopping between $xy,zx,yz$ orbitals, respectively.
When both $\theta_i$ and $\theta_j$ do not equal $b_{ij}$ we find $\hat h(\theta_i,\theta_j)=0$, 
whereas either one of them equals $b_{ij}$, the ferromagnetic interaction $\hat h(\theta_i,\theta_j)=\hat h_1$
is realized. If both equal $b_{ij}$, we find an antiferromagnetic interaction, $\hat h_2$.
\par
Since states with different orbital and lattice configurations do not mix quantum mechanically, 
Eq.(\ref{eq:finalmodel}) is block diagonalized into these configuration sectors, 
and the full quantum many-body treatment is to solve the quantum $S=1$ model 
interacting via a bond network of given orbital and lattice configurations. 
The ground state is determined by comparing the lowest energies of different orbital-lattice configurations. 
\par
The straightforward simplification is given by replacing $\hat h_\ell$ with the constant energy value. 
To make a proper choice of the energy constant values, 
we evaluate the quantum energy of the trimer state in Eq.(\ref{eq:finalmodel}) 
realized in a finite size cluster by ED, (see Appendix A), 
in comparison with the sum of constant bond energies realized on a same size cluster. 
If we consider an isolated $S=1$ pair, we find $\langle {\bm S}_i\cdot {\bm S}_j \rangle=-2$ for $S_{\rm tot}=0$ 
and $\langle {\bm S}_i\cdot {\bm S}_j \rangle=1$ for $S_{\rm tot}=2$. 
However, substituting the former to $\langle \hat h_2\rangle$ 
overestimates the quantum energy gain expected for the trimer state on the bulk triangular lattice. 
We thus adopt the classical value $\langle {\bm S}_i\cdot {\bm S}_j \rangle=\mp 1$, 
which gives 
\begin{align}
& \langle \hat h_1\rangle = -J_F-J_{F0}, \notag \\
& \langle \hat h_2\rangle = -2J_{AF}. 
\label{eq:jvalues}
\end{align}
Apart from these values, this kind of treatment is the same as the discussion made by Pen, {\it et al.}, 
where they adopt $ \langle \hat h_1\rangle=-J_0$ and  $\langle \hat h_2\rangle=-2J_0$. 
The introduction of lattice displacements in our case largely modifies the energetics of orbital configurations. 
For the regular types of lattice distortion, we provide in Table I 
the number of bonds per site units shown in Fig.~\ref{f4}, together with
the number of unchanged, short, and long bonds, 
each with four different types of orbital configurations.
The energies of the orbital-lattice model are evaluated classically by counting the number of these bonds 
for the given types of orbital and lattice configurations. 
For large $\eta$ where the Hund's coupling is relevant, 
$J_{AF}$ decreases and $J_F$ increases, which favors magnetically ordered states 
rather than singlets, and the system becomes more classical and deviates from the 
orbital-spin-lattice model that treats the longer-range quantum correlations 
between $S=1$'s.

\begin{figure*}[t]
    \centering
    \includegraphics[width=\textwidth]{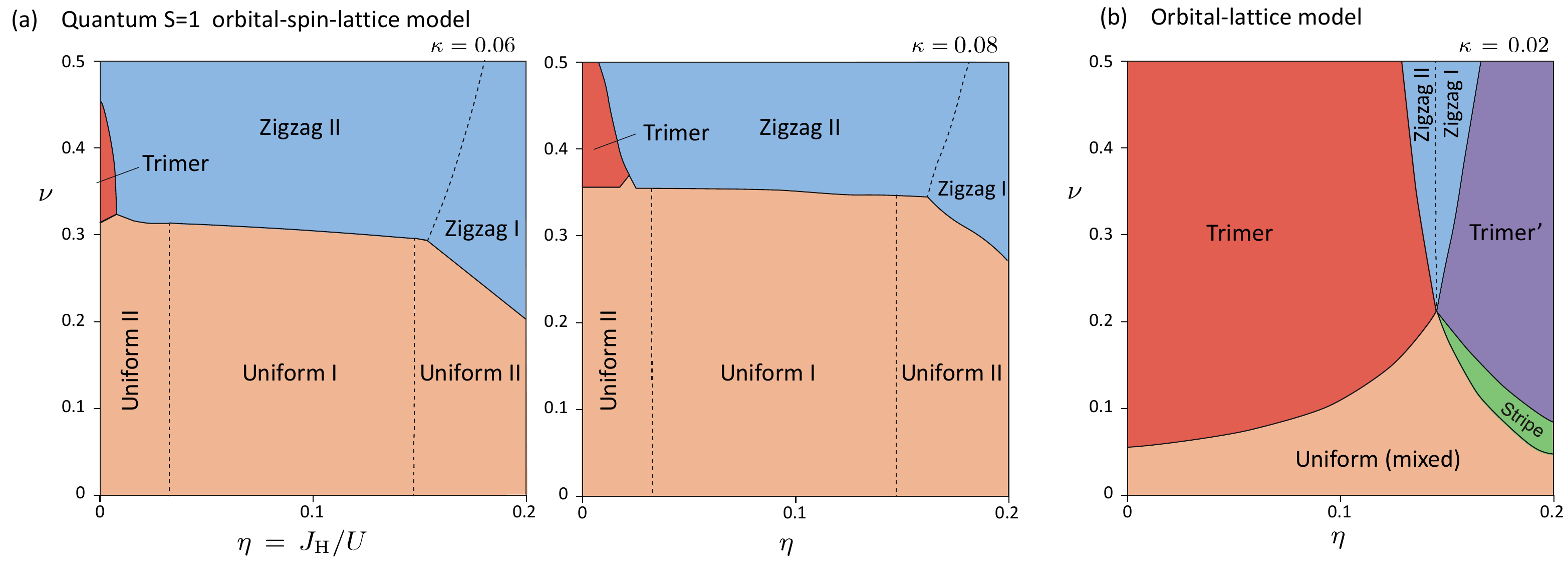}
    \caption{Ground state phase diagrams as functions of the Hund's coupling strength
    $\eta=J_{\rm H}/U$ and magnetoelastic coupling $\nu$ for 
    (a) the orbital-spin-lattice model in Eq.(\ref{eq:finalmodel}) with $\kappa=0.06, 0.08$ where 
    quantum $S=1$ sectors are solved by ED. 
    (b) The orbital-lattice model adopting Eq.(\ref{eq:jvalues}) for the bond-energy evaluation with $\kappa=0.02$.
     We find Trimer, Zigzag, and Uniform phases, 
     where the Trimer has a three-spin singlet state, and 
     Zigzag and Uniform phases have Ferrimagnetic ordering of types I, II, or their mixture. 
     There are additional phases labeled Trimer$'$ (elongated trimer) and stripe in (b) at large $\eta$.} 
    \label{f5}
\end{figure*}
%

\section{Numerical results}
\label{sec3}
\subsection{$T=0$ ground states}
We first investigate the ground state of the orbital-spin-lattice model in Eq.(\ref{eq:finalmodel})
by taking full account of the quantum mechanical $S=1$ degrees of freedom using ED. 
Although there are $(3\times 6)^N$ different orbital-lattice configurations for system size $N$,
most of the nonperiodic ones can generally be excluded from the ground state.
We thus confine ourselves to the orbital configurations giving $J_{AF/F}$ shown in Fig.~\ref{f4} (a)--(d),
combined with their preferable lattice displacements shown in Fig.~\ref{f4} (e)--(g).
By diagonalizing the $S=1$ Hamiltonian with interaction networks determined by the orbital-lattice configuration, 
we obtain the lowest energy and compare it with those of other configurations, 
and identify the ground state as the one with the lowest total energy. 
In the following we set $J_0=t^2/U=1$ as a unit energy, 
and implitly use $\kappa$ and temperature $T$ in that unit. 
\par
The ED calculations are performed on a $N=6\times2$ cluster shown in Fig.~\ref{f4},
where the unit cell is chosen appropriately so that both the orbital configuration and the lattice distortion
are compatible with the periodic boundary conditions. 
The phase diagrams on the plane of $\eta=J_H/U$ and $\nu$ obtained from our ED results are presented in Fig.~\ref{f5}(a) for
two different choices of $\kappa=0.06$ and $0.08$. 
The key finding is the realization of the Trimer phase in the small range of $\eta$
with substantial lattice distortions, $\nu\sim 0.3-0.45$.
As $\eta$ increases, the system undergoes a phase transition to one of the Zigzag phases.
The Zigzag I (II) phase corresponds to a combination of the zigzag displacement pattern 
[Fig.~\ref{f4} (g)] and the Ferri I (II) state.
The reason why $\eta$ works to stabilize the Ferri phase against the Trimer singlet phase is 
because the Hund's rule favors the two electrons to form a more robust triplet, 
which is reflected in the increase/decrease of spin exchange couplings, $J_F$/$J_{AF}$. 
For $\nu\lesssim 0.3$, the lattices are undistorted while the orbital configuration is either Ferri I or Ferri II.
We denote these phases as Uniform I and Uniform II. 
\par
By comparing the phase diagrams with two different $\kappa$'s, we find that
the increase of elastic energy loss with $\kappa$ pushes up the
distorted phases (Trimer and Zigzag) to larger $\nu$, as they need more energy gain in terms of $t$ to compete with $\kappa$.
At the same time, without the lattice distortion,
the three-spin singlet phase has relatively small quantum $S=1$ energy gain compared to the Ferri phase,
although the numbers of FM and AFM bonds are both $2N$ and $N$ (see Table I),
namely there is a substantial longer range correlation for the latter.
When including the lattice distortion, this spatial coherence is suppressed,
and thus the advantage of the latter becomes less.
Accordingly, the Trimer phase with three-spin singlet extends to $\eta\sim 0.03$ for $\kappa=0.08$.
\par
Next, we show in Fig.~\ref{f5} (b) the phase diagram 
of the classical orbital-lattice model that adopts Eq.(\ref{eq:jvalues}) for the evaluation of bond energy. 
The ground state energies of each phase are calculated analytically and are shown in Appendix A.
Although we set $\kappa=0.02$, which is smaller than that for the quantum orbital-spin-lattice model, 
the Trimer phase is extended over $\eta\lesssim 0.16$ when $\nu$ is sufficiently large. 
With increasing $\eta$, this trimer phase is replaced successively by the Zigzag II and Zigzag I phases. 
This behavior is consistent with the tendency found in the ED phase diagram in Fig.~\ref{f5}(a). 
The trimer phase is overestimated compared to the quantum case, which is because 
the Zigzag phases should naturally have large quantum energy gain due to long-distance spin-spin fluctuations, 
compared to the trimer phase consisting of local singlets, and the former gain is discarded in the classical orbital-lattice model. 
\par
On the other hand, when $\eta$ is decreased, the system enters the Uniform phase, 
in which all lattice displacements are ordered in a particular direction. 
As the relative distance between lattice sites is the same as the undistorted lattice, 
this ordering is purely statistical; The lattice degrees of freedom form a $q=6$ clock model which is known 
to exhibit ``ferro'' type of long-range order at the lowest temperature
\cite{Tobochnik1982PRB,Challa1986PRB,Okabe2002PRB,Chatterjee2018PRE,Li2020PRE}. 
The orbital ordering becomes the mixed component of Ferri I and Ferri II, as the energies of these two are equal in the classical case. 
For $\eta\gtrsim 0.16$, we also find another trimer phase (Trimer$'$, which has elongated trimer bonds) and a stripe phase. 
However, these phases are the artifacts of the model, since the present classical approach is 
not reliable in the large-$\eta$ regime. 
\par
The comparison between the phase diagrams of the two types of models shows that 
although the orbital-lattice model overestimates the stability of the Trimer phase 
combined with the lattice degrees of freedom, both cases provide the same tendency;
First, without lattice distortions ($\nu=0$),
there is no room to stabilize the Trimer singlet.
Second, there is an onset $\nu$ to have the Trimer phase, while this phase is replaced by the Zigzag phase with Ferri order
at large $\eta$ as the $J_{F}$ gradually increases and $J_{AF}$ decreases. 
Therefore, by properly tuning the value of $\kappa$ to be smaller
and by taking account of the fact that Trimer is overestimated and extends to much larger $\eta$,
one can make use of the orbital-lattice model to study the overall
property of the competition between the trimer and the other magnetically ordered phases. 
\par
Let us discuss the relevance of the present model parameters to the material trend in layered LiV$X_2$ ($X$=O,S).
Experimentally, increasing the chalcogen ionic radius increases both the effective hopping $t$
and the itinerant character of the electronic state. 
Indeed, LiVO$_2$ remains insulating, LiVS$_2$ becomes metallic at the higher temperature range above the trimer formation, 
and LiVSe$_2$ stays metallic without developing the low-temperature trimer phase \cite{Katayama2024,Kojima2023PRB}.
In the language of the present strong-coupling model, this increase in $t$ enhances the electronic
energy scale $J_0=t^2/U$, which relatively decreases the elastic constant $\kappa$ in our phase diagram. 
This makes the electronic energy gain dominant over the elastic energy cost. 
Resultantly, Zigzag state can overtake the Trimer, which consistently explains 
the overall tendency observed experimentally\cite{Katayama2024,Katayama2021npjQM,Kojima2023PRB}.


\subsection{Thermodynamic properties}
Based on the above considerations,
we adopt the orbital-lattice model as an effective model to
study the thermodynamic properties of the system.
As the model includes only the classical degrees of freedom,
the Monte Carlo simulations based on the Metropolis algorithm can be applied \cite{Metropolis1953JCP}.
One Monte Carlo step (MCS) consists of attempting local Metropolis updates of
orbital and lattice degrees of freedom at all sites. 
By dividing the system into three sublattice groups, 
we update the orbital and lattice variables independently within each group 
which enables efficient parallelization. 
\par
Equilibrium simulations are performed on triangular lattices on $N=L\times L$ rhombuses 
with $L=144$ under periodic boundary conditions 
with the fixed elastic constant, $\kappa=0.02$. 
After the equilibration performed at $T=10J_0$, the system is gradually cooled down by performing
$100\,000$ MCSs at each temperature point until it reaches the lowest temperature. 
Then, the system is heated along the same temperature path.
At each temperature, the first $50\,000$ MCSs are 
discarded, and the measurements are taken for every 100 MCSs.
The specific heat is evaluated from the energy fluctuation as
$C=(\langle E^2\rangle-\langle E\rangle^2)/N T^2$.
\begin{figure*}[ht]
    \centering
    \includegraphics[width=\linewidth]{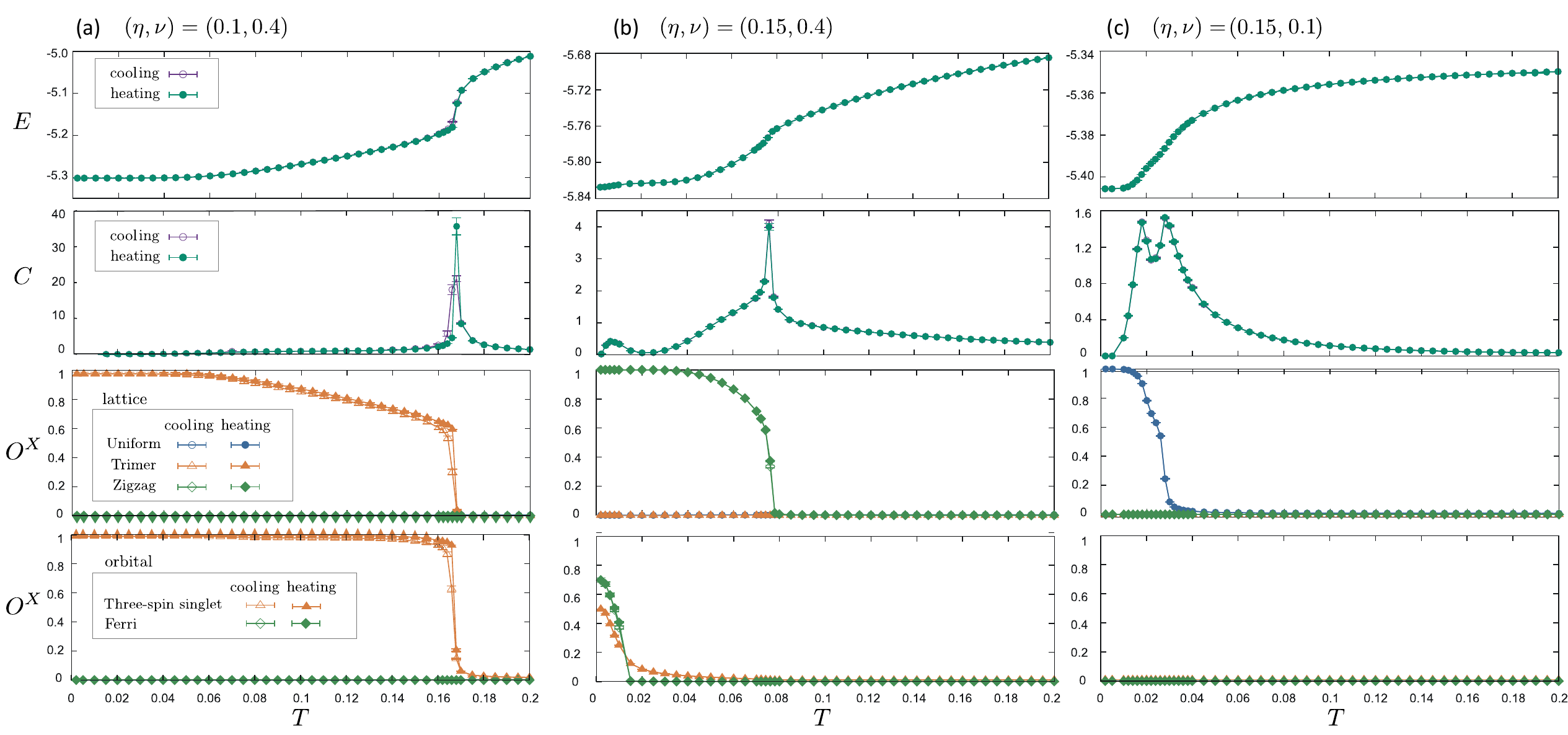}
    \caption{
    Temperature dependence of the energy $E$, specific heat $C$, both per site, and lattice and orbital
    order parameters $O^X$ obtained by equilibrium MC simulations on a $144\times 144$ size cluster. 
    The data are obtained separately through the cooling and heating processes. 
    Standard error bars are shown for each data point evaluated over 144 independent MC runs. 
    We choose the parameters, (a) $(\eta,\nu)=(0.1,0.4)$ with the Trimer ground state, 
    (b) $(0.15,0.4)$ with the Zigzag ground state, and (c) $(0.15,0.1)$ with the Uniform (mixed) ground state. 
    }
    \label{f6}
\end{figure*}
\begin{figure}[tbp]
    \centering
    \includegraphics[width=0.8\linewidth]{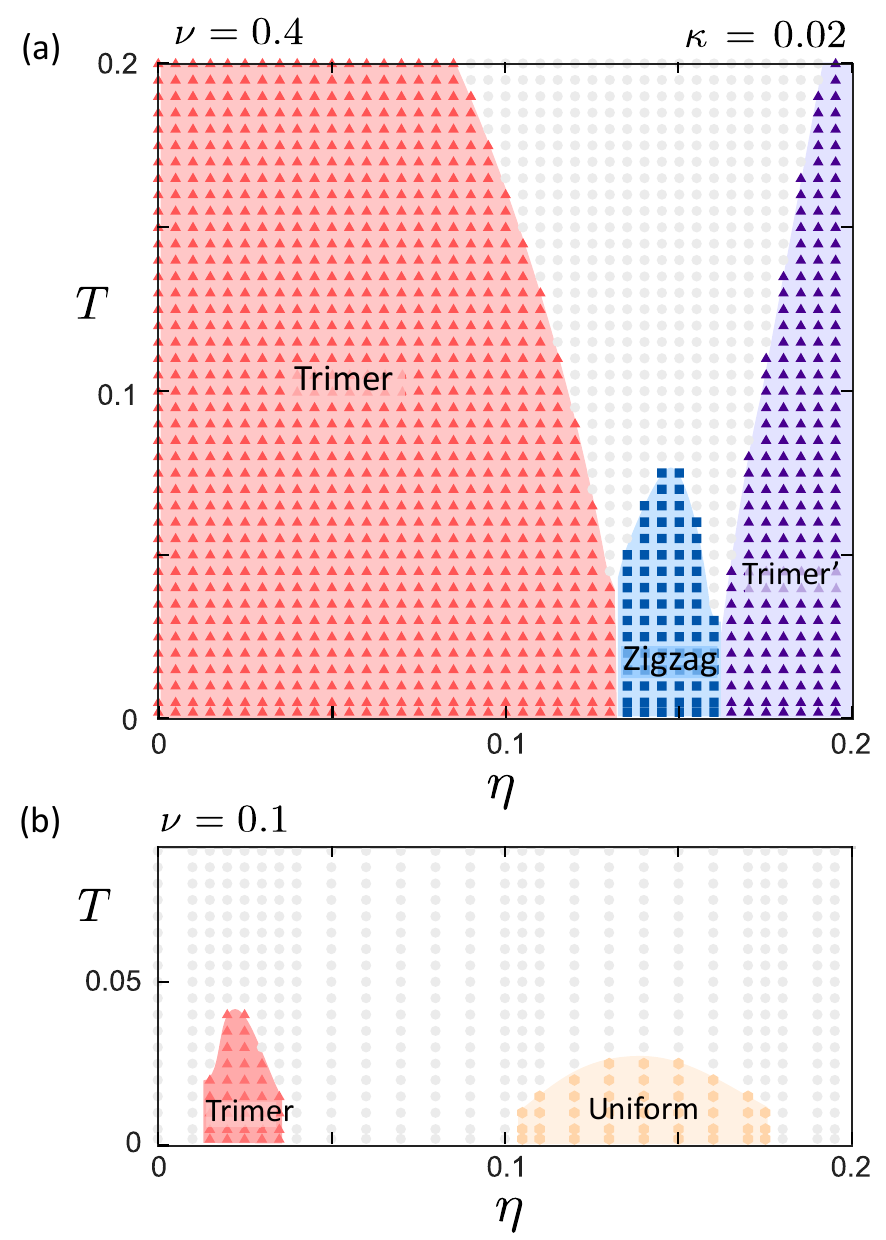}
    \caption{
    Finite-temperature phase diagrams of the orbital-lattice model obtained by
    equilibrium MC simulations for (a) $\nu=0.4$ and (b) $\nu=0.1$. 
    Six independent MC runs are performed for each data point over both the heating and cooling processes. 
    The phase labels indicate the types of orders that have $O^X \ge 0.5$. 
    }
    \label{f7}
\end{figure}
\par
To identify the orbital and lattice orderings, 
we encode the local orbital and lattice configurations by complex clock variables as 
\begin{align}
    X_\ell^{\rm orb}=e^{i2\pi j_\ell^{\rm orb}/3}, \qquad
    X_\ell^{\rm lat}=e^{i\pi j_\ell^{\rm lat}/3},
\end{align}
where $j_\ell^{\rm orb}=0,1,2$ labels the three orbital states and 
$j_\ell^{\rm lat}=0,\cdots,5$ labels the six displacement directions.
For $X_\ell=X_\ell^{\rm orb}$ or $X_\ell^{\rm lat}$, we introduce the structure factor
\begin{align}
    S_X({\bm q})
    =\frac{1}{N}\left|\sum_\ell X_\ell e^{i{\bm q}\cdot{\bm r}_\ell}\right|^2
    =\frac{1}{N}\sum_{\ell,m} X_\ell^* X_m
    e^{i{\bm q}\cdot({\bm r}_m-{\bm r}_\ell)} .
\end{align}
The uniform, trimer, and zigzag components are then measured by
\begin{align}
    & O_{\rm uniform}^{X} = S_X(\Gamma), \\
    & O_{\rm trimer}^{X} = \left|S_X(K_1)-S_X(K_2)\right|, \\
    & O_{\rm zigzag}^{X} =
    \left|S_X(M_1)+\omega S_X(M_2)+\omega^2 S_X(M_3)\right|,
\end{align}
respectively, where $\omega=e^{2\pi i/3}$, 
and the high-symmetry ${\bm k}$ points are 
$\Gamma=(0,0)$, 
$K_1=(1,\frac{1}{\sqrt{3}})$, 
$K_2=(-1,\frac{1}{\sqrt{3}})$, 
$M_1=(1,0)$, 
$M_2=(\frac{1}{2},\frac{\sqrt{3}}{2})$, and 
$M_3=(\frac{3}{2},\frac{\sqrt{3}}{2})$. 
The absolute values are taken before averaging over MC samples so that 
different symmetry-related domains do not cancel each other. 
\par
The representative temperature dependences of 
the energy, specific heat, and lattice and orbital order parameters are shown in 
Figs.~\ref{f6}(a)-(c) 
for the parameter values of $(\eta,\nu)=(0.1,0.4)$, $(0.15,0.4)$, and $(0.15,0.1)$. 
\par
We first discuss the case of $(\eta,\nu)=(0.1,0.4)$. 
The most important feature is the hysteretic behavior associated with the
trimer ordering. 
The energy and the trimer order parameters show abrupt changes at nearly the
same temperature, indicating the first-order transition.
The specific heat exhibits a divergent behavior and a sharp drop consistent with the first-order character. 
The specific heat and order parameters show hysteretic behavior between heating and cooling 
processes, which also supports the first-order transition. 
\par
In the case of  $(\eta,\nu)=(0.15,0.4)$, the system enters 
a Zigzag phase via second order transition, as one can see from the 
continuous growth of the lattice order parameter at around $T\simeq 0.08$. 
There is another transition at $T\simeq 0.016$ associated with the orbital ordering, 
which is also a continuous transition. 
We note that although the orbital trimer order parameter is nonzero at lower temperature, 
it does not indicate the trimer ordering because the mixture of nonzero Zigzag orders 
running in different directions yields 
finite trimer order parameters, which is the artifact of the definition. 
\par
Finally, we discuss the case of $(\eta,\nu)=(0.15,0.1)$.
Here, continuous transitions associated with the lattice degree of freedom is 
observed at around $T\simeq 0.02-0.03$, which exhibit two peaks in the specific heat. 
The lattice displacement takes place in the same direction for all sites, 
that yields a peak at $S_X(\Gamma)$. 
Physically, this indicates an undistorted lattice, and may seem to have nothing to do with the lattice ordering. 
However, if we focus solely on the lattice degrees of freedom, the model reduces 
to the $q=6$ clock model which is known to show two-peaks indicating the 
Kosterlitz-Thouless (KT) transitions\cite{Tobochnik1982PRB,Challa1986PRB,Okabe2002PRB,Chatterjee2018PRE,Li2020PRE}; 
the higher temperature peak indicates the transition to the quasi-long range ordered 
$xy$-like phase, and the lower peak marks the transition to the uniformly ordered phase. 
The energy shows a two-step-like behavior characteristic of such transitions. 
In that context, these transitions naturally occur because we assumed the discrete clock-type of 
lattice distortion, and has nothing to do with the original materials. 
Nevertheless, this transition supports a static uniform lattice structure, allowing for the 
orbital degrees of freedom to choose the lowest energy state of their own. 
The resultant Zigzag I and II phases are energetically degenerate, 
and appear as their mixture, which is the reason why we cannot find nonzero contributions 
from the orbital order parameters. 
\par
In Fig.~\ref{f7}, we show the finite temperature phase diagrams along the $\eta$ axis 
for two different choices of parameters, $\nu=0.4$ and $0.1$. 
At $\nu=0.4$, the Trimer phase is quite robust and sustains up to $T\sim 0.3$, 
in contrast to the Zigzag phase that forms a dome at low temperature. 
At small $\nu$, the Trimer phase is strongly suppressed, and so as the Uniform mixed phase, 
and the ordered phase only appears below the lowest temperature in our calculation, $\sim 0.02$.

\section{Summary}
In this work, we investigated the microscopic origin of trimer formation in orbitally degenerate transition-metal compounds by deriving and analyzing effective low-energy models based on a multiorbital Kanamori-Hubbard Hamiltonian on a triangular lattice. 
Motivated by LiVS$_2$ and LiVO$_2$, where each V$^{3+}$ ion hosts two electrons in the threefold-degenerate $t_{2g}$ orbitals, we focused on the strong-coupling regime in which Hund's coupling stabilizes local $S=1$ triplets formed by electrons occupying different orbitals.

Projecting the microscopic Hamiltonian onto the local $S=1$ manifold, we derived an orbital-spin model in which the geometry of the spin-exchange network is determined by the orbital configuration. To account for experimentally observed structural distortions, we further introduced lattice degrees of freedom through local ionic displacements. The relative displacements between neighboring ions modify bond lengths, thereby changing the exchange couplings and competing with the associated elastic energy cost. The resulting orbital-spin-lattice model captures the cooperative interplay among spin, orbital, and lattice degrees of freedom.

The ground-state phase diagram obtained by exact diagonalization of the quantum $S=1$ orbital-spin-lattice model reveals strong competition among Trimer, Zigzag, and Ferrimagnetic phases. 
We confirmed that the orbital-selective exchange interactions alone are insufficient to stabilize the Trimer phase. 
A robust three-spin singlet phase of $S=1$ emerges only when a substantial Trimer lattice distortions are allowed 
in the presence of a substantial magnetoelastic coupling, 
that enhance the exchange anisotropy between shortened and elongated bonds. 
In the absence of such distortions, the system favors the Ferri phase, while Zigzag ordering becomes competitive in distorted lattices.
This phase competition also provides a natural interpretation of the material trend in LiV$X_2$; 
where the increase of chalcogen ionic radius enhances the effective hopping $t$ and hence the
electronic energy scale $J_0=t^2/U$, making the elastic energy scale $\kappa$ relatively smaller. 
As a result, the Trimer state is replaced by the Zigzag state, 
consistent with the observation of slow dynamics related to the zigzag geometry when 
$X$ changes from O to S\cite{Katayama2021npjQM}.
\par
To access larger system sizes and finite-temperature properties, we further introduced a simplified orbital-lattice model in which the quantum $S=1$ degrees of freedom are integrated out and replaced by effective AF and F bond energies. Despite quantitatively overestimating the stability of the Trimer phase due to the neglect of long-range quantum spin correlations, the simplified model reproduces the essential structure of the ground-state phase diagram and provides a useful framework for studying thermal phase transitions.

Monte Carlo simulations of the orbital-lattice model revealed a rich finite-temperature phase diagram. The Trimer phase 
appears through a first-order transition accompanied by simultaneous ordering of lattice and orbital degrees of freedom and exhibits a pronounced hysteresis between cooling and heating processes. The Zigzag phase develops through a continuous transition, with lattice and orbital ordering occurring at distinct temperatures. For weaker lattice coupling, the Ferri phase is stabilized through a Kosterlitz-Thouless transition from an intermediate quasi-long-range ordered regime. This behavior originates from the effective six-state clock-like character of the lattice sector associated with the undistorted configuration.

Our results demonstrate that lattice distortions qualitatively reshape the energetics of orbitally degenerate triangular-lattice systems. Previous studies based solely on spin-orbital interactions concluded that the Trimer state is unstable against Ferrimagnetic ordering, while a fully quantum-mechanical treatment of the competing phases remained largely unexplored. By systematically comparing the orbital-spin-lattice and orbital-lattice descriptions, we identify lattice-modulated orbital-selective exchange interactions as the key ingredient responsible for stabilizing trimerization. 
The present work therefore establishes a microscopic framework for understanding cluster formation 
in frustrated transition-metal compounds 
and highlights the crucial role of coupled spin, orbital, and lattice degrees of freedom in determining their collective states.

\begin{acknowledgments}
We are grateful for fruitful discussions with Atsushi Ikeda, Naoyuki Katayama, Keita Kojima, and Yuta Sakai. 
The work is supported by the Grant-in-Aid for Transformative Research Areas A “Extreme Universe” (KAKENHI Grant No. JP21H05191) 
and other KAKENHI Grants No. JP21K03440, JP26H00635,JP26K00627. 
The numerical calculations are performed using the facilities of the Supercomputer Center, 
the Institute for Solid State Physics, the University of Tokyo. 
\end{acknowledgments}
\appendix
\section{Comparison of energies of the two models}
\label{sec:gsene}
\begin{figure}[tbp]
    \centering
    \includegraphics[width=\linewidth]{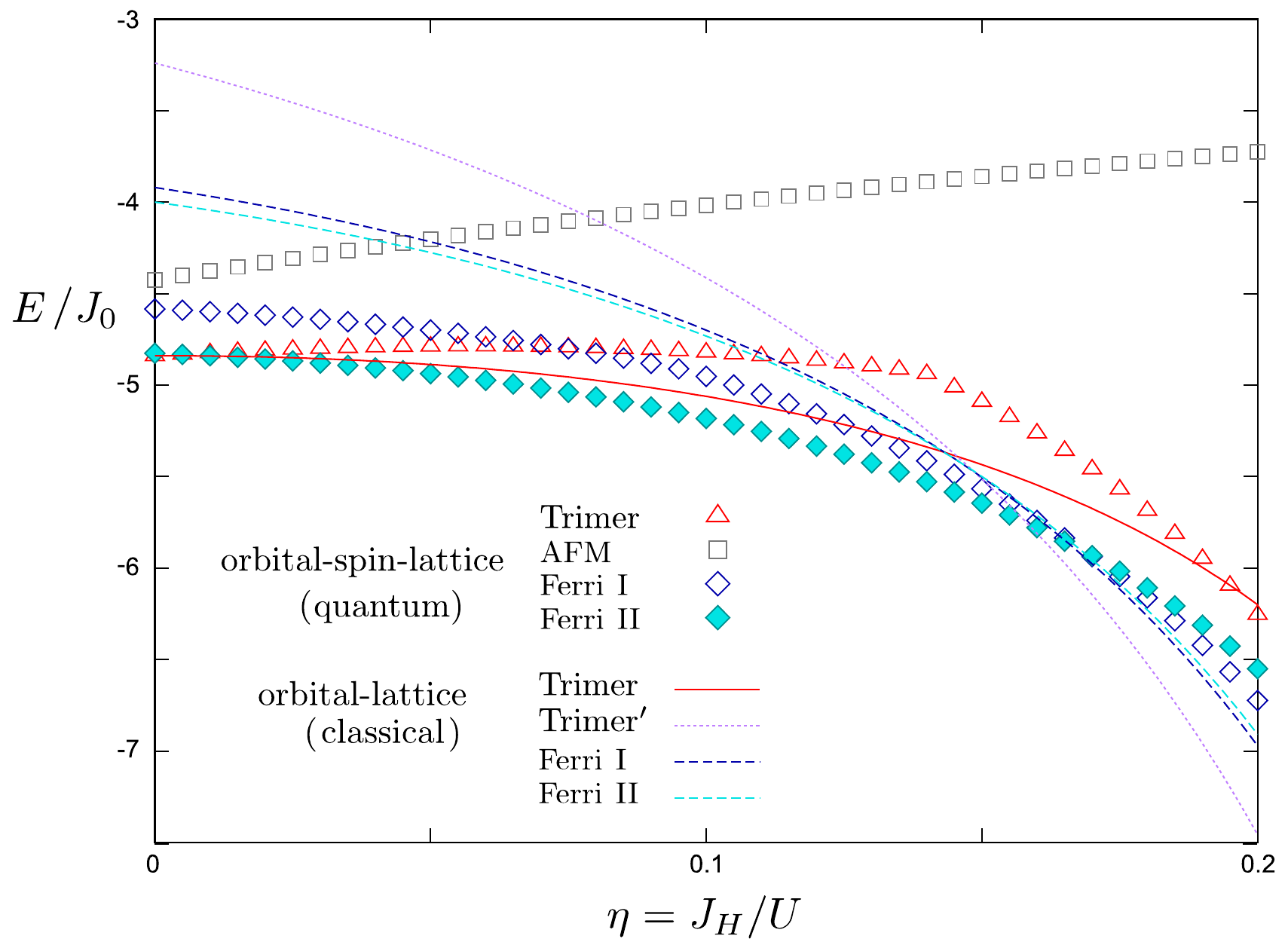}
    \caption{
    Energies of the Trimer, Ferri I, II, and AFM (square geometry) states of the orbital-spin-lattice model calculated by ED 
    using the cluster shown in Fig.~\ref{f4}, plotted as functions of $\eta$. 
    Solid and broken lines are the Trimer/Trimer$'$ and Ferri I, II phases realized in the classical orbital-lattice model 
    for the same parameter, to be compared with the quantum case. 
    }
    \label{fapp}
\end{figure}

To make a quantitative comparison between the quantum $S=1$ orbital-spin-lattice model and the classical orbital-lattice model, 
we show in Fig.~\ref{fapp} the energies of the representative phases for $\nu=0.4$ and $\kappa=0.06$. 
The Trimer phase is energetically almost degenerate with the Ferri phases in the quantum ED results, 
and the Ferri phases become lower in energy when we introduce $\eta$, 
and becomes the ground states we mentioned as Uniform I/II phases in the phase diagram. 
The introduction of the lattice-displacement degrees of freedom 
allows the substantial energy gain of the Trimer state, and provides a finite window to become a ground state at small $\eta$. 
\par
The energies of the orbital-lattice model are evaluated by adopting Eq.(\ref{eq:jvalues}) as the AF and F bond energies. 
There, we chose $\langle \bm S_i\cdot \bm S_j\rangle=-1$ and $1$ for $\hat h_2$ and $\hat h_1$, respectively, 
which is because this choice allows the Trimer state energy to be almost equal to the one evaluated for the quantum $S=1$ case
(see Fig.~\ref{fapp}). 
\par
The total energies of the representative phases evaluated for Eq.(\ref{eq:jvalues}) are given as
\begin{align}
&E_{\rm Trimer}=- \frac{2(1-\nu/2)^2}{1-3\eta} - \frac{2(1+\eta)}{1+2\eta}(1+\nu)^2 + 6\kappa, \notag \\
&E_{\rm Trimer'}= -\frac{2(1+\nu/2)^2}{1-3\eta} - \frac{2(1+\eta)}{1+2\eta}(1-\nu)^2 + 6\kappa, \notag \\
&E_{\rm Ferri I} 
    = -\frac{(1+\eta)(2+\nu^2)}{1+2\eta} - \frac{2+3\nu^2/2}{1-3\eta} +8\kappa,  \notag \\
&E_{\rm Ferri II} 
    = -\frac{(1+\eta)(2+2\nu^2)}{1+2\eta} - \frac{2-\nu^2}{1-3\eta} + 8\kappa, \notag \\
&E_{\rm Uniform} = -\frac{2}{1-3\eta} - \frac{2(1+\eta)}{1+2\eta}, \notag \\
&E_{\rm Stripe} = -\frac{1+(1+\nu/2)^2}{1-3\eta} - \frac{2(1+\eta)(1-\nu/2)^2}{1+2\eta} + 2\kappa.
\end{align}
The overall $\eta$ dependence of these energy values is consistent with the quantum cases, 
although the crossing points differ and modify the phase diagram, 
as we discussed in the main text. 

\bibliography{livs2}
\end{document}